\documentclass[article, 12pt]{elsarticle}
\usepackage[utf8]{inputenc}

\setcitestyle{authoryear}
\setcitestyle{citesep={;}}

\usepackage{lineno,hyperref}
\modulolinenumbers[5]

\journal{arXiv}
 
\usepackage{tikz}
\usetikzlibrary{shapes.geometric, arrows}
\tikzstyle{process} = [rectangle, minimum width=3cm, minimum height=1cm, text centered, draw=black, fill=white!30]
\tikzstyle{arrow} = [thick,->,>=stealth]

\usepackage{hyperref}
\hypersetup{
    colorlinks=true,
    linkcolor=blue,
    filecolor=magenta,      
    urlcolor=cyan,
    pdftitle={Surrogate Nerve Model}}
\usepackage{amsmath}
\usepackage{amssymb}
\usepackage{dsfont} 
\usepackage{mathtools}
\usepackage{algpseudocode}
\usepackage{float}
\usepackage{graphicx}
\usepackage{wrapfig}
\usepackage{makecell}
\usepackage{mathrsfs}
\usepackage{multirow}

\newcommand{\wave}{\phi_e(t)}
\newcommand{\pot}{\psi_e^f}
\newcommand{\thres}{\hat I_e^f}

\begin{document}
\begin{frontmatter}
\title{Optimizing Stimulus Energy for Cochlear Implants with a Machine Learning Model of the Auditory Nerve}

\author[liacs]{Jacob de Nobel}
\author[liacs]{Anna V. Kononova}
\author[lumc]{Jeroen Briaire}
\author{Johan Frijns$^{\textrm{b, c}}$}
\author[liacs]{Thomas B\"ack}

\address[liacs]{Leiden Institute of Advanced Computer Science, Niels Bohrweg 1, Leiden, Netherlands}
\address[lumc]{Department of Otorhinolaryngology, Leiden University Medical Center, Albinusdreef 2, Leiden, Netherlands}
\address[cog]{Leiden Institute for Brain and Cognition, Wassenaarseweg 52, Leiden, Netherlands}

\begin{abstract}
Performing simulations with a realistic biophysical auditory nerve fiber model can be very time consuming, due to the complex nature of the calculations involved. Here, a surrogate (approximate) model of such an auditory nerve fiber model was developed using machine learning methods, to perform simulations more efficiently. Several machine learning models were compared, of which a Convolutional Neural Network showed the best performance. In fact, the Convolutional Neural Network was able to emulate the behavior of the auditory nerve fiber model with extremely high similarity ($R^2 > 0.99$), tested under a wide range of experimental conditions, whilst reducing the simulation time by five orders of magnitude. In addition, we introduce a method for randomly generating charge-balanced waveforms using hyperplane projection. In the second part of this paper, the Convolutional Neural Network surrogate model was used by an Evolutionary Algorithm to optimize the shape of the stimulus waveform in terms energy efficiency. The resulting waveforms resemble a positive Gaussian-like peak, preceded by an elongated negative phase.  When comparing the energy of the waveforms generated by the Evolutionary Algorithm with the commonly used square wave, energy decreases of 8\% - 45\% were observed for different pulse durations. These results were validated with the original auditory nerve fiber model, which demonstrates that our proposed surrogate model can be used as its accurate and efficient replacement.
\end{abstract}

\begin{keyword}
Cochlear Implants, Optimization, Auditory Nerve, Neural Model, Machine Learning, Evolutionary Algorithms
\end{keyword}

\end{frontmatter}

\section{Introduction} 
\label{sec:int}
Cochlear Implants (CI) are neuroprostheses which use direct electrical stimulation of the auditory nerve to restore a sense of sound and speech understanding to people with profound sensorineural hearing loss. A condition, which is estimated to affect roughly 5\% of the global population~\citep{olusanya2014global}. These devices consist of an external part containing a battery, a microphone, and a sound processor, which transforms sounds into coded signals that are sent via a wireless transmitter to an implanted component. The implanted electronics include a receiver and a stimulator that converts the coded signals to a pattern of electrical pulses that stimulate the auditory nerve via a multi-channel electrode array in the cochlea (inner ear). \\

The external part of the device has to be worn on the side of the head, behind the ear, and the social stigma~\citep{Rapporte058406} that is associated with wearing a hearing aid can be a potential barrier for use. In fact, the size and visibility of hearing aids has been identified as one of the main factors limiting adoption~\citep{sizeproblem}. For the CI, a large part of the external unit is taken up by the battery, and thus if we could decrease the implant's power consumption, this would lead to smaller batteries and, consequently, smaller devices. \\

In this work, we focus on neural stimulation of the auditory nerve, which has been reported to account for up to 90\% of the total power consumed by the device~\citep{ciga}. Specifically, we analyze the shape of the waveforms used to stimulate the auditory nerve fiber (ANF), whilst optimizing energy efficiency. In practice, the CI is often configured to use square biphasic waveforms, even though many studies show the potential benefit of using alternate shapes. For example, the introduction of an inter-phase gap~\citep{CARLYON2005210} or the use of asymmetric pulses~\citep{Macherey2006} has been shown to lower the stimulation threshold. Moreover, it has been shown that the human auditory system is more sensitive to the anodic phase of biphasic stimuli~\citep{macherey2008higher, undurraga2010polarity}. In some studies~\citep{jezernik2005energy, sahin2007non}, the reduced threshold energy of using exponentially decaying/increasing waveforms has been reported. \citet{Wongsarnpigoon2010energy} investigates square, exponential and linear ramp waveforms for monophasic stimulation, but finds that none of these waveforms are simultaneously energy, power and charge optimal. In a later study~\citep{ga2010}, an Evolutionary Algorithm (described as a Genetic Algorithm (GA) by the authors) is used to optimize waveform shape for energy and finds that a truncated Gaussian pulse followed by a rectangular charge balancing phase is best. This methodology was used in~\citet{yip2017energy} specifically for an ANF and found that a biphasic pulse with exponentially decaying cathodic phase followed by a rectangular anodic phase provides the lowest energy at threshold level. \\

In order to investigate the effect of electrical stimulation on the auditory nerve cells in the cochlea, simulation with computer models can be used~\citep{Frijns1995, Frijns2000IntegratedUO, 3dmodelsreview}. This can be an attractive alternative for collecting measurements on CI, as \textit{in vivo} experiments require human subjects, and a significant effort to conduct. Early work in computational modelling with compartment models yielded the activating function~\citep{rattay1986, warman1993}, which is proportional to the second-order spatial derivative of the extracellular potential along the axon, and can be used to approximate the influence of electrical stimulation on neural excitation. Contemporary models~\citep{rattay2001model, smit2010threshold, kalkman2022} combine 3D modelling techniques with non-linear multi-node ANF models, and conceptually split up the simulation task into two processes: I) The calculation of the electrical field potentials in the geometry of the cochlea, and II) the simulation of the response of a nerve fiber to an externally applied potential field~\citep{briairechapter4}. The increased complexity of these later models comes at a substantial computational cost, as they require the solution to large set of differential equations to accurately simulate fiber behavior. \\

Surrogate modelling~\citep{surrogategs, jin2011surrogate} is a technique used in engineering and optimization that uses an approximate model of a process for which the outcome is not easily measured or computed. This has the potential to massively speed up calculation time, as the evaluation of a surrogate is often orders of magnitude faster than the original process. In this work, we use a surrogate modelling approach to create a statistical approximate for the second of the two simulation processes of the model presented in~\citet{kalkman2022}, the ANF model~\citep{briaire2005unraveling}. With the surrogate, it becomes possible to do a much larger number of simulations than was previously possible with the original model, which allows researchers to perform experiments on a much larger scale. \\ 

This paper features two contributions. Firstly, five different machine learning models are compared in detail when tasked with constructing a surrogate for the active nerve model used in~\citet{kalkman2022}. This results in favor of a Convolutional Neural Network (CNN), which predicts the neural response with very high accuracy. Secondly, we draw inspiration from~\citet{ga2010} and~\citet{yip2017energy} and use the surrogate to optimize the stimulus waveform for energy with an Evolutionary Algorithm (EA). The surrogate effectively enables this type of analysis, as the large number of evaluations required to do the optimization would have been infeasible to do with the original model. We propose a new method for generating charge-balanced waveforms, using hyperplane projection. This allows us to generate any charge-balanced waveform, whilst in previous works the shapes that could be generated were limited. This experiment results in a potential alternative to the square biphasic pulses used in practice and indicates substantial energy savings. \\

The paper is organized as follows; first, an overview of the used methods, models and data is given in Section~\ref{sec:method}. Section~\ref{sec:modsel} includes the comparison of different machine learning models for training the surrogate. The waveform optimization experiment can be found in Section~\ref{sec:optimization}, which is followed by a conclusion to the paper in Section~\ref{sec:conclusion}.

\section{Methods}
\label{sec:method}
In this section, we will first introduce the model of electrical hearing used in this study, and explain the two components of the model, the 3D Volume Conduction model (Section~\ref{sec:volc}) and the ANF model (Section~\ref{sec:nerve}). This is followed by a description of the design process of a surrogate for the latter of these two models, including a detailed description of the data gathering process (Section~\ref{sec:data}) and machine learning techniques used (Section~\ref{sec:regr}).

\subsection{Model of Electrical Hearing}
\label{sec:models}
In this research, we build on~\citet{kalkman2022}, where an updated version of a computational model of the implanted human cochlea is presented. As mentioned in the introduction, this model consists of two underlying models, which will both be described in more detail here. 

\subsubsection{3D Volume Conduction Model}
\label{sec:volc}
\begin{figure}[h]
    \centering
    \includegraphics[width=.8\textwidth]{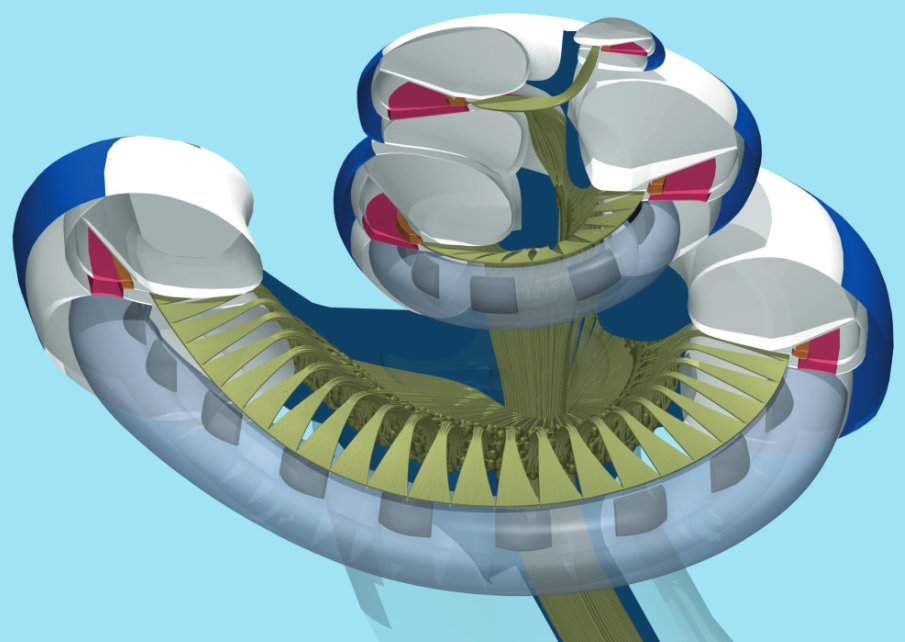}
    \caption{Three-dimensional cut-through of a cochlea with a laterally implanted electrode array, as used by the volume conduction model. Nerve fibers are shown in yellow, and the electrodes contacts can be identified in grey/black. Figure taken from~\citet{kalkman2022}.}
    \label{fig:cochlea}
\end{figure}

The volume conduction part of the model, uses a boundary element model (BEM) of the cochlea that can model arbitrary cochlear geometries, implanted with a multi-channel electrode array~\citep{Briaire2008Chapter3}. An example of such a geometry can be seen in Figure~\ref{fig:cochlea}. These geometries are characterized by a 3D mesh of polygons, which represent the different areas of the cochlea. When current is injected into the inner ear by the electrodes, this induces a potential distribution. The calculation of this distribution, termed the volume conduction problem, consists of finding the solution to the Poisson equation:
\begin{equation}
    \nabla^2 \psi = - \frac{I}{\sigma},
\end{equation}
where $\psi$ represents the potential distribution, $I$ the injected current, and $\sigma$ the conductivity of the medium. Because the cochlea does not consist of a heterogeneous medium (i.e., different tissue types have different levels of conductivity), every polygon in the mesh is associated with the levels of conductivity $\sigma$ corresponding to the tissues it separates. In the model, fibers are represented by several cylindrical segments, for which at each segment the extracellular potential is calculated. This yields a vector of ($\approx$ 115) regularly spaced measurements of the electrical potential along the fiber. This vector is then normalized over the input current, which produces a transfer function with its values in k$\Omega$. For simplicity, we refer to this transfer function by the term potential distribution in the remainder of the paper. The potential distribution induced by electrode $e$ for a nerve fiber $f$ is denoted with $\pot$. As can be read in \ref{sec:pot}, we can approximate this ($\approx$ 115)-valued vector by a fifth order polynomial, which allows us to represent $\pot$ as a vector in $\mathbb{R}^5$. 

\subsubsection{Active Model of the Auditory Nerve}
\label{sec:nerve}
The second part of the computational model architecture, consists of an auditory nerve fiber model~\citep{dekker2014, kalkman2022} with human Schwarz-Reid-Bostock (SRB) kinetics~\citep{Schwarz1995}, and calculates the neural response elicited by a stimulus. It is a non-linear double cable model developed to display fundamental mammalian nerve fiber properties, such as spike conduction velocity and repetitive firing behavior~\citep{Frijns2000IntegratedUO}. The model represents a human bipolar High Spontaneous Rate (HSR) fiber, which is the most prevalent type of primary auditory nerve fiber~\citep{Frijns2000IntegratedUO}, and is modelled in accordance with known human data~\citep{briairechapter7}.  \\

The ANF model $\mathbf{f}$ takes an arbitrary stimulus waveform $\wave$, which is a discrete univariate time series of current $I$ over a given experimental time window. Based on the potential distribution $\pot$ provided by the 3D volume conduction model (see Section \ref{sec:volc}) and a stimulus waveform $\wave$, the model calculates a threshold level $\thres$:
\begin{equation}
    \mathbf{f}(\mathbf{x}) =  \thres \in \mathbb{R}, 
\end{equation} where $\mathbf{x}:=(\pot, \wave)$. 
$\thres$ represents the relative amount of current that must be given by an electrode $e$ using a waveform $\wave$ for a fiber $f$ to spike. In other words, if the waveform $\wave$ is scaled by $\thres$, an action potential can be observed. The computation of $\thres$ requires the solution to a system of differential equations. This is solved using a classic backwards Euler method, which takes a considerable amount of time to compute, and grows linearly with the length of $\wave$. 

\subsection{Surrogate Model of the Auditory Nerve Fiber}
\label{sec:surro}
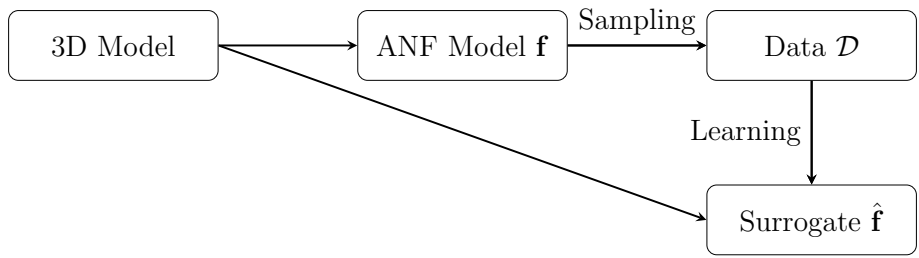
\begin{figure}[h]
\centering
\resizebox{0.9\textwidth}{!}{
\begin{tikzpicture}[node distance=5cm]
\tikzstyle{comp2} = [draw, rectangle, rounded corners, minimum height=1cm, minimum width=3cm]

\node (vm) [comp2]{3D Model};
\node (am) [comp2,  right of=vm]{ANF Model $\mathbf{f}$};
\node (db) [comp2,  right of=am]{Data $\mathcal{D}$};
\node (sm) [comp2,  below of=db, yshift=2.5cm]{Surrogate $\hat{\mathbf{f}}$};

\draw [arrow] (vm) -- (am);
\draw [arrow] (vm.east) -- (sm.west);
\draw [arrow] (am) -- (db);
\draw [arrow] (db) -- (sm);
\draw [arrow] (am) -- node[anchor=south] {Sampling} (db);
\draw [arrow] (db) -- node[anchor=east] {Learning} (sm);
\end{tikzpicture}
}
\caption{Schematic overview of the design of a surrogate model for the ANF model For a detailed description on the structure of the sampled data, see Section~\ref{sec:data}.} 
\label{fig:sdia}
\end{figure}

The surrogate model is intended as a substitution of the ANF model described in Section~\ref{sec:nerve}, in order to reduce its significant computation time. The goal is to construct a statistical approximate $\hat{\mathbf{f}}$ of the nerve model $\mathbf{f}$, through the use of machine learning. By presenting the surrogate with a large number of training examples generated with the nerve model, the surrogate can be trained to function as its replacement. A schematic of this process is shown in Figure~\ref{fig:sdia}.

\subsubsection{Regression}
\label{sec:regr}
Designing a surrogate for the ANF model can be postulated as a regression task, modeling the relation between the independent model parameters $\mathcal{X}$ and the dependent model outcomes $\mathbf{y}$. Formally, we define the data set for training the surrogate model as a set of input-target pairs $\mathcal{D} \coloneqq \{(\mathbf{x}_i \in \mathbb{R}^{85}, y_i \in \mathbb{R})_{i=1, \dots, n}\}$, where each input $\mathbf{x}_i$ consists of two entities: the potential distribution $\pot$ computed by the volume conduction model and the stimulus waveform $\wave$. The target value $y_i$, equals the activation threshold $ \thres$. \\

We define a regression model as a function $\hat{\mathbf{f}}_\theta(\mathcal{X}) \rightarrow \hat{\mathbf{y}}$ parameterized by $\theta$, which outputs a predicted value $\hat{\mathbf{y}}$, representing activation thresholds ($\thres$), for a given data set $\mathcal{X}$ of nerve model simulation parameters $(\pot, \wave)$. By providing the model with a set of training examples, we search for the regression model parameter set $\theta$, which cause it to predict a $\hat{\mathbf{y}}$ with a minimal error compared to the true value $\mathbf{y}$. We evaluate this using the coefficient of determination ($R^2$), which is used to indicate the 'goodness of fit' for a regression model. Defined on $[-\infty, 1]$, $R^2$ denotes the proportion of the dependent variable ($\mathbf{y}$) which is correctly inferred from the independent variables ($\mathcal{X}$): 
\begin{equation}
    R^2(\mathbf{y}, \hat{\mathbf{y}}) = 1 - \frac{\sum_{i=1}^n (\mathbf{y}_i - \hat{\mathbf{y}}_i)^2}{\sum_{i=1}^n (\mathbf{y}_i - \bar{\mathbf{y}})^2}
\end{equation} Here, $\bar{\mathbf{y}}$ denotes the average value of $\mathbf{y}$. Note that a constant model which predicts the expected value for $\mathbf{y}$ yields an $R^2$ of 0, and a perfect model receives a value of 1. $R^2$ is calculated for the training, and test data sets separately, see also Section~\ref{sec:crossval}. 

\subsubsection{Regression models}
As was already mentioned in the introduction (Section~\ref{sec:int}), we consider five machine learning models for the surrogate (regression) model: 
\begin{enumerate} 
    \item \textbf{Polynomial Elastic Net (PEN)} is a polynomial regression model, that combines the $\mathscr{L}1$ and $\mathscr{L}2$ regularization factors of the lasso and ridge regression methods. Regularization reduces model complexity, effectively performing an in-place feature selection on the input data by penalizing the regression model for having large weights. This produces a sparse model, which can be especially useful when many of the input features are correlated~\citep{elastic}. 
    \item \textbf{Random Forests (RF)} ~\citep{randomforests} is a form of ensemble learning that uses multiple decision trees, utilizing the principle that many weak classifiers together form a strong classifier, through a process called bagging. In bagging (also known as bootstrap-aggregating), $b$ samples of $\mathcal{X}$ are taken with replacement, and on each of these subsets $\mathcal{X}_b$ a decision tree $\mathcal{T}_b$ is trained. Predictions are made through a form of majority voting, where the average of the assigned outputs of all $\mathcal{T}_b$ is used for the final prediction. 
    \item \textbf{Gradient Tree Boosting (GB)} is another type of ensemble learning which uses an ensemble of decision trees, similar to RF. In contrast to RF, the decision trees are created additively, where each tree is built to improve upon the already existing ones. This is achieved via gradient descent of a differentiable loss function specified by the user. GB often performs better than RF, but can be prone to overfitting, which is why regularization methods have been introduced~\citep{gradboost}. 
    \item \textbf{Multi-Layer Perceptron (MLP)} is a fully connected feed-forward artificial neural network~\citep{goodfellow2016deep}. It consists of at least three layers of perceptrons or nodes, an input layer, several hidden layers and an output layer. Each node in the network has a linear activation function that maps the weighted inputs to a scaled output: $z(\mathbf{w} \cdot \mathbf{x}_i) \mapsto \mathbb{R}$, which is then passed to the next layer of the network. The weights $\mathbf{w}$ of each node are obtained via the backpropagation of error terms in a form of gradient descent on the complete network. In regression, the output layer of the model, which consists of a single node, is evaluated without an activation function, yielding a continuous value. 
    \item \textbf{Convolutional Neural Network (CNN)} is a deep neural network, which sets itself apart from regular neural networks by the fact that they use a convolution instead of a matrix multiplication when processing data~\citep{goodfellow2016deep}. This allows CNN's to be highly effective at processing data with an inherent structure, such as images and audio signals, as they are invariant to the shifts to the input. The structure of the network is similar to a MLP, except that is has at least one convolutional layer as one of their hidden layers. The convolution operation features a kernel that slides over the input and generates a feature map, which is passed as the input to the next layer. By assembling these feature maps, the model can learn increasingly complex patterns and abstract features from the data, taking advantage of their hierarchical structure.  
\end{enumerate}
For PEN, RF and the MLP, we used the implementations as provided by the $\textsc{scikit-Learn}$~\citep{sklearn} (version 0.24.2) Python library. For the gradient boosting algorithm, the implementation from the library $\textsc{xgboost}$~\citep{xgboost} (version 1.5.2) was used, and for the CNN, we used the $\textsc{keras}$~\citep{keras} API for  $\textsc{tensorflow}$~\citep{tf} (version 2.6.2). All experiments were conducted using Python 3.7.5. 

\subsection{Data}
\label{sec:data}
Data was collected by performing simulations with the ANF model described in Section~\ref{sec:nerve}, for various experimental conditions. Our research data includes potential distributions $\pot$ (see Section~\ref{sec:volc}) for five different cochlear geometries~\citep{kalkman2022}, implanted with both lateral wall (LW) and perimodiolar (PM) electrode arrays, which are model equivalents of a HiFocus1J cochlear implant with 16 electrode contacts. Samples are collected for up to 3\,200 nerve fibers (placed uniformly along three cochlear turns) with three levels of neural degeneration: i) healthy fibers (H), ii) 'short-terminal' (ST) fibers and iii) fibers with completely degenerated (CD) dendrites~\citep{Kalkman2015a}. \\
\begin{table}[h]
    \centering
    \begin{tabular}{l|c|c|c}
        &  $\mathbf{A}$ & $\mathbf{B}$ & $\mathbf{C}$  \\ \hline \hline
    \textbf{electrode type} & LW, PM & LW & LW \\ 
    \textbf{fiber health} & H, ST, CD  & H, ST, CD &  H \\
    \textbf{\# cochleae} & 5 & 1 & 1 \\
    \textbf{\# electrodes} & 16 & 1 & 1 \\
    \textbf{\# fibers} & 3\,200 & 3\,200 & 1 \\
    \textbf{\# pulses} & 1 & 1\,296 & 48\,000 \\ 
    \textbf{\# samples} & 1\,466\,189 & 12\,441\,600 & 48\,000 \\ \hline
    \end{tabular}
    \caption{High level overview of the  data sets used in this work; $\mathbf{A}, \mathbf{B}$ and $\mathbf{C}$.}
    \label{tab:datasets}
\end{table}

In this work, we consider three data sets, for which an overview is given in Table~\ref{tab:datasets}. As can been seen from the table, each data set has different components, which allows us to investigate the parameters of the ANF model from contrasting aspects. Data set $\mathbf{A}$ has the highest variability in potential distributions $\pot$, and includes data for several cochlear geometries and electrodes, but only considers a single stimulus waveform $\wave$. A much larger number of stimulus waveforms are included in data set $\mathbf{B}$ for a single geometry and electrode contact. In contrast, data set $\mathbf{C}$ only varies the stimulus waveform and keeps all other parameters constant. More detail on the stimulus waveforms used is given in Sections \ref{sec:stim} and \ref{sec:lhs}. \\ 

Note that the number of samples for data set $\mathbf{A}$ is not a perfect multiple of the number of experimental conditions used to generate it. This is because simulations with parameters that were not able to yield a finite value for $\thres$ were excluded, as the maximum value for $\thres$ was limited to 100. Additionally, all data was scaled to zero mean and unit variance prior to training the surrogate model. For every feature (i.e. column) $\mathbf{x}^{i}$ in our data, using the mean $\bar{\mathbf{x}}^{i}$ and standard deviation $SD(\mathbf{x}^{i})$ of the training data set (see Section~\ref{sec:crossval} for more details on train/test data sets), the standardized feature values were computed as follows:
\begin{equation}
  \mathbf{x}^{i} = \frac{\mathbf{x}^{i}- \bar{\mathbf{x}}^{i}}{SD(\mathbf{x}^{i})}.
\end{equation}

\subsubsection{Stimulus}
\label{sec:stim}
In practice, a stimulus waveform $\wave$ is a univariate real-valued time series with a fixed sampling rate, representing the active current $I$ in ampere ($A$) of an electrode $e$ over a sampling window. Here, a time window of 160$\mu s$ is considered with a sampling rate of 2$\mu s$. This allows us to represent a stimulus waveform as a vector in $\mathbb{R}^{80}$, with each component bounded on $[-0.01, 0.01]$. \\ 

\begin{figure}[tb] 
    \centering
    \includegraphics[width=\textwidth]{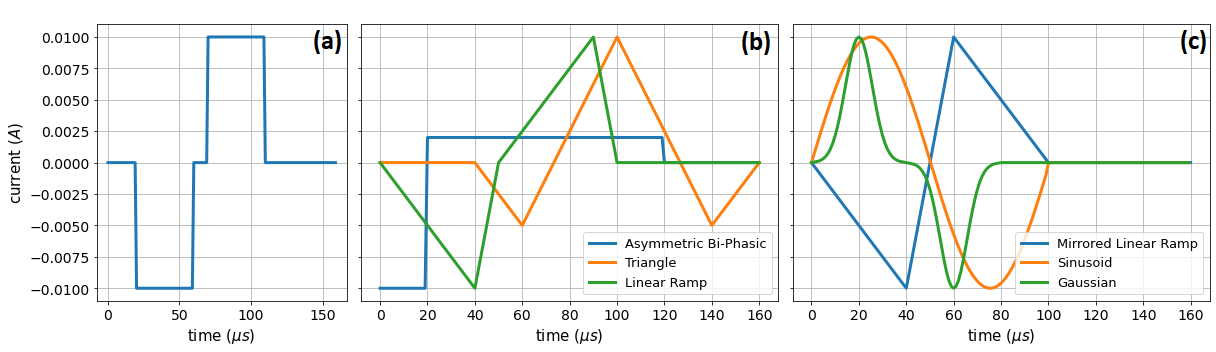}
    \caption{Examples of the predefined waveforms used in this paper.}
    \label{fig:b_pulses}
\end{figure}

In standard operating conditions, a cochlear implant is often configured to produce square biphasic pulses~\citep{guinpig_pulses,Macherey2006,ga2010,yip2017energy}, which can be adjusted parameterically in terms of the phase-duration, amplitude and polarity. An example of a cathodic-first biphasic pulse, with a phase duration of $50\mu s$, a delay of $20\mu s$ and an inter-phase gap of $10 \mu s$ is shown in Figure~\ref{fig:b_pulses}\textbf{(a)}. Nevertheless, the cochlear implant can be configured to transmit arbitrary stimulus waveforms, given that they are charge balanced, i.e.:
 
\begin{equation}
\label{eq:const}
  \sum_{t=0}^\infty \wave = 0 
\end{equation}
This is to ensure that the implant does not cause tissue damage due to an unbalanced charge, which can introduce gases, corrosion products and pH changes as byproduct that can be detrimental to the user~\citep{huang1999electrical}. \\ 
 
In this work, all simulations are conducted using monopolar stimulation~\citep{kalkman2022}. Data is generated for a single cathodic-first biphasic pulse, with a phase-duration of 18$\mu s$, no inter-phase gap and no offset. For this pulse, data was collected for all the cochlear geometries and electrode configurations. We denote this data set $\mathbf{A}$ (see Table~\ref{tab:datasets}). \\

Additionally, data was generated for pulses found commonly in the literature (an overview of these can be found in Figure~\ref{fig:b_pulses} and Table~\ref{tab:pulses}), which are included in data set $\mathbf{B}$. For each of these pulses, variations were generated by scaling them to a different time duration, in $2 \mu s$ intervals, up to a length of 160 $\mu s$, using three different methods:
\begin{itemize}
    \item \textbf{Gap-Insertion} ($\mathcal{GI})$ takes the base pulse and inserts an inter-phase gap at all zero-crossings of the pulse.
    \item \textbf{Extension} ($\mathcal{E}$) grows the pulse by copying every element, i.e: $$(x_1, \dots, x_n) \mapsto (x_1, x_1, \dots, x_n, x_n).$$
    \item \textbf{Interpolation} ($\mathcal{I}$) computes the linear interpolation of the pulse at $2n$ points, and replaces the original with the interpolation, i.e: $$(x_1, \dots, x_n) \mapsto (x_1^{a}, x_1^{b}, \dots, x_n^{a}, x_n^{b}),$$ where $x_i^{a}$ and $x_i^{b}$ denote the interpolated points around point $x_i$.
\end{itemize}

Starting with a base pulse $x$ of length $n$, the $\mathcal{E}$ and $\mathcal{I}$ operators recursively increase the duration of the pulse up to the sampling window (160 $\mu s$). Using these three operators, 648 different stimulus waveforms were generated. Each of these waveforms was then also considered with reversed polarity by multiplying it by -1, yielding a total of 1\,296 pulses/waveforms. Simulations with the ANF model for these pulses (i.e. data set $\mathbf{B}$) were only conducted for the HC3A geometry, and one electrode (electrode 9, i.e., in the middle of the array) due to the significant amount of time in running these simulations (several weeks on a 64 machine compute cluster).

\begin{table}[tb] 
\centering
\begin{tabular}{l|l|l|l}
\textbf{Name}        & \textbf{Example}           & \textbf{Extension method}   & \textbf{Proportion}                                  \\ \hline \hline
Biphasic            & Fig.~\ref{fig:b_pulses}\textbf{(a) }  &   $\mathcal{GI}$,  $\mathcal{E}$,  $\mathcal{I}$ &         0.247                    \\
Asymmetric biphasic & Fig.~\ref{fig:b_pulses}\textbf{(b)}    &   $\mathcal{GI}$,  $\mathcal{E}$,  $\mathcal{I}$ &        0.517                   \\
Triangle             & Fig.~\ref{fig:b_pulses}\textbf{(b)}        &   $\mathcal{GI}$,  $\mathcal{E}$,  $\mathcal{I}$ &    0.097                    \\
Linear ramp          & Fig.~\ref{fig:b_pulses}\textbf{(b)}         &   $\mathcal{E}$,  $\mathcal{I}$ &                    0.052                    \\
Mirrored Linear ramp & Fig.~\ref{fig:b_pulses}\textbf{(c)} &   $\mathcal{E}$,  $\mathcal{I}$ &                   0.046                    \\
Sinusoid             & Fig.~\ref{fig:b_pulses}\textbf{(c)} & $\mathcal{I}$ &                        0.009                    \\
Gaussian             & Fig.~\ref{fig:b_pulses}\textbf{(c)}      &   $\mathcal{I}$ &                                       0.032                    \\ \hline
\end{tabular}
\caption{An overview of the stimulus waveforms considered in the experiments for data set $\mathbf{B}$, and the methods used to generate their variants. The proportion of samples in the data for a given base is also shown.}
\label{tab:pulses}
\end{table}

\subsubsection{Generating charge-balanced waveforms}
\label{sec:projection}
\begin{figure}[h]
    \centering
    \includegraphics[width=.55\textwidth]{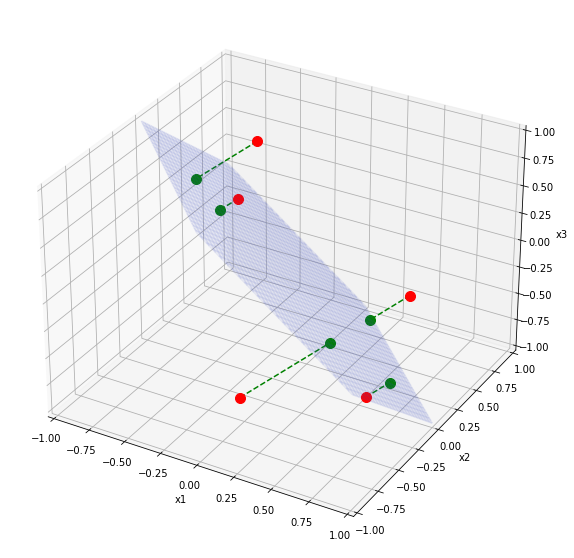}
    \caption{Visualization of the charge-balancing transformation (eq.~\ref{eq:chargebal}), in three dimensions. The non-violating hyperplane $H$ is shown in blue, the original samples $\mathbf{x}$ in red and the transformed samples $\mathbf{x}'$ in green. The dashed line indicates the applied transformation on each sample.}
    \label{fig:chargebal}
\end{figure}

As mentioned in Section~\ref{sec:stim}, the stimulus waveform is represented by real-valued vector from $[-0.01, 0.01]^{80}$. In order to satisfy the charge-balancing constraint (eq.~\ref{eq:const}), we can project every vector in a sample to the non-violating hyperplane: 
\begin{equation}
    H:= \{\mathbf{x}: \langle \mathbf{x}, \mathbf{1} \rangle = b\},
\end{equation} where $b = 0$. The inner product between $\mathbf{x}$ and $\mathbf{1}$ is denoted by $\langle \mathbf{x}, \mathbf{1} \rangle$, and $\mathbf{1}$ represents a vector with all ones of the same size as $\mathbf{x}$. We then project every sampled vector $\mathbf{x}$ onto $H$ as follows:
\begin{equation}
    \label{eq:chargebal}
    \mathbf{x}' = \mathbf{x} - \frac{\langle \mathbf{x}, \mathbf{1} \rangle - b}{|| \mathbf{1} ||^2}  \mathbf{1}
\end{equation}
This ensures that every projected vector $\mathbf{x}'$ becomes charge-balanced (with b=0) and can be used as a stimulus waveform $\wave$. Figure~\ref{fig:chargebal} provides a visual intuition of this transformation for three dimensional samples. From the figure, we can see that for every point in the three dimensional space, we can compute an orthogonal transformation onto $H$, which satisfies eq.~\ref{eq:const}.

\subsubsection{Latin Hypercube Sampling}
\label{sec:lhs}
Since the distribution of stimulus waveforms used in data sets $\mathbf{A}$ and $\mathbf{B}$ have a relatively low variance, we conduct a Latin Hypercube Sampling (LHS) of $\wave$ to get a broader range of training examples for the surrogate model. We label this data set $\mathbf{C}$ (see Table~\ref{tab:datasets}). \\

Latin Hypercube sampling is a technique for generating quasi-random samples from a continuous multidimensional distribution, which ensures the drawn samples are distributed evenly in the sample space~\citep{lhs}. It involves dividing the cumulative density function (CDF) for each of the dimensions into $N$ evenly spaced partitions, which are used to construct a multidimensional grid or hypercube. Random samples are then taken in such a way that only one sample is drawn for each axis-aligned hyperplane. For example, in a two-dimensional grid, this means that for every row and every column, there is only one sample~\citep{lhs2}. This produces a set of robustly stratified random samples, which, in contrast to traditional random sampling, provides the assumption that the samples reflect the true variability of the underlying distribution.  \\

As it is computationally infeasible to simulate a large number of LH samples using the ANF model for all geometries and nerve fibers, we only do this for a single nerve fiber and electrode combination. In this way, we only vary the stimulus waveform $\wave$ in our data, while the potential distribution $\pot$ remains fixed. We thus only generate samples for a single nerve fiber, but since the $\wave$ and $\pot$ are independent model parameters, we can expect that observations made from this data set will generalize to all fibers. We use the potential distribution induced by electrode 9 on fiber 1213, which was chosen because this fiber has the lowest average threshold $ \thres$ for electrode 9, which is also included in data set $\mathbf{B}$. \\

For this data set $\mathbf{C}$, we draw a fixed number (8\,000) of random samples for stimulus waveforms $\wave$ of length 80. Since we also want to include shorter waveforms in our sample, we draw an additional fixed number (2\,000) of samples for each length in $\{2, 6,\dots, 74, 78\}$. Each of these samples is then padded with with zeros on the right in order to produce a vector in $\mathbb{R}^{80}$. This yields a total of 48\,000 vectors in our sample for data set $\mathbf{C}$. Note that we cannot directly use these vectors as stimulus waveforms, since because they are generated in a pseudo-random fashion, they do not satisfy constraint~(eq. \ref{eq:const}). However, by applying the transformation described in Section~\ref{sec:projection}, we charge-balance each of the 48\,000 random samples in data set $\mathbf{C}$. 

\section{Model selection}
\label{sec:modsel} 
To establish which of the investigated machine learning models (see Section~\ref{sec:regr}) is best suited for surrogate construction, we conduct two different experiments, discussed in more detail in the following sections. Firstly, we conduct a hyperparameter tuning experiment on a small subset of the data set $\mathbf{A}$, to find good (hyper) parameter sets for each of the investigated models. Then, using these parameter sets, we evaluate model performance in a large-scale cross-validation experiment on all the data sets ($\mathbf{A}, \mathbf{B}, \mathbf{C}$) separately. All experiments in this section are performed on a Linux server running CentOS 7, with 256 CPU's and 500 GB of RAM.

\subsection{Hyperparameter tuning}
\label{sec:hyp}
In order to determine the best settings for each of the aforementioned machine learning models (see Section~\ref{sec:regr}), we conduct a hyperparameter tuning experiment, using a grid search on their parameters. A grid search involves an iterative procedure that performs an exhaustive search over all combinations of parameter values specified by the user and returns the parameters which yield the highest performance. In order to get a robust estimate of the performance associated with a given parameter set, we use 3-fold cross-validation (see Section~\ref{sec:crossval} for more details on cross-validation). This requires a large number of fitting operations for each of the models and would be very costly if evaluated on one of the complete data sets, due to their relatively large size. Thus, we construct a down-sampled grid-search data set $\mathcal{D}_g$ from 1\% of the samples in data set $\mathbf{A}$, taken uniformly at random, which includes 14\,661 samples. We hold back 33\% of this data for testing ($\mathcal{D}_g^{test}$) and perform the grid search on the remaining 67\% ($\mathcal{D}_g^{train}$). Then for each model, the performance of the best parameter set found by the grid search is validated by testing on $\mathcal{D}_g^{test}$, using a model trained on $\mathcal{D}_g^{train}$. \\

\begin{table}[tb] 
\centering
\begin{tabular}{l|l|l}
  \textbf{Model}       &    \textbf{Parameter} &  \textbf{Best value}  \\ \hline \hline
  \multirow{3}{*}{\makecell{\textbf{Polynomial}  \\ \textbf{Elastic Net}}} & $\textsc{degree}$ & 3 \\ & $\alpha$ & 0.1 \\ & $\mathcal{L}\textsc{1-ratio}$ & 0.1 \\   \hline
  \multirow{2}{*}{\makecell{\textbf{Multi-Layer}\\ \textbf{Perceptron}}} & $\textsc{activation}$ & $\textsc{tanh}$ \\& $\textsc{hidden\_layer\_sizes}$ & (500, 250, 100)  \\ \hline
  \multirow{5}{*}{\makecell{\textbf{Convolutional} \\ \textbf{Neural Network}}} & $\textsc{n\_filters}$ & 512\\ &$\textsc{kernel\_size}$ & 16\\ & $\textsc{pool\_size}$ & 4 \\ & $\textsc{dropout\_rate}$ & 0.2\\ & $\textsc{n\_nodes\_dense}$ & 100  \\ \hline
  \multirow{6}{*}{\makecell{\textbf{Gradient Tree} \\ \textbf{Boosting}}}&  $\gamma$ & 0.8\\&  $\textsc{learning\_rate}$ & 0.1 \\& $\textsc{max\_depth}$ & 14\\& $\textsc{reg}_\alpha$ & 0.1 \\& $\textsc{reg}_\lambda$ & 12.8\\& $\textsc{tweedie\_variance\_power}$ & 1.2 \\ \hline
  \multirow{4}{*}{\makecell{\textbf{Random}  \\ \textbf{Forests}}} & $\textsc{n\_estimators}$ & 500\\ & $\textsc{max\_depth}$ & 30\\ & $\textsc{min\_samples\_split}$ & 2\\ & $\textsc{min\_samples\_leaf}$ & 1 \\ \hline
  \end{tabular}
\caption{Best values of the parameters of machine learning models investigated, found by the grid search.}
\label{tab:grids}
\end{table}

\begin{figure}[tb] 
    \centering
    \includegraphics[width=\textwidth]{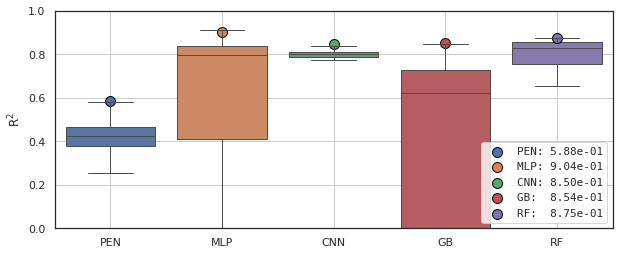}
    \caption{Grid search results. Each boxplot shows the distribution of the mean value of $R^2$ for every searched parameter set over a 3-fold cross-validation on the training data set $\mathcal{D}_g^{train}$. In addition, the validation score attained on $\mathcal{D}_g^{test}$ for the best-found parameter set for each of the models is shown as a circle, the precise values of these dots are given in the legend. Note that the y-axis is clipped at 0.}
    \label{fig:grids}
\end{figure}

Table~\ref{tab:grids} summarizes the list of searched parameters and their best values. For more detail on the parameters, we refer the interested reader to the $\textsc{scikit-learn}$, $\textsc{tensorflow}$ and $\textsc{xdgboost}$ documentation~\citep{sklearn,xgboost, keras, tf}. \\

In Figure~\ref{fig:grids}, the results of the grid search experiments are shown. The figure shows two kinds of results. Firstly, it shows the distribution of the 3-fold cross-validation $R^2$ scores on $\mathcal{D}_g^{train}$ graphed as box plots for parameter sets of each considered model. Secondly, the $R^2$ score attained by each model when using the best parameter set on $\mathcal{D}_g^{test}$ is shown in the figure as circles. The precise value ($R^2$) for each dot is shown in the legend. The Figure clarifies that the best parameter sets for all models show similar performance on this data set ($\mathcal{D}_g$), except for the PEN model, which seems to be less suited for this particular regression problem. Since PEN is the least complex model, this is to be expected. The highest score on the test set ($0.904$) was attained by the MLP model.

\subsection{Validation}
\label{sec:crossval}
In this section, we evaluate the hyperparameters sets which showed the highest performance in the grid-search experiment (see Table~\ref{tab:grids}). Each model, with its best hyperparameter setting, is evaluated on the full $\mathbf{A}$, $\mathbf{B}$ and $\mathbf{C}$ data sets (see Table~\ref{tab:datasets}) in this experiment. This shows the true performance of each model on each data set. Moreover, since each of the three data sets is generated with different experimental conditions, performance recorded on $\mathcal{D}_g$ might not generalize to the other data sets. In order to assess how the results for each model will generalize to unseen data, we can use \textit{cross-validation}~\citep{crossval}, which is a re-sampling method that uses different partitions of the data to iteratively train and test a machine learning model. The goal of cross-validation is to evaluate the performance of a model on data that was not used to train it, providing insight into how it generalizes to an independent data set. \\

\paragraph{10-fold cross-validation}
Here, we perform a 10-fold cross-validation, meaning we partition each data set in 10 equally sized groups or so-called folds, selected uniformly at random. Then we sequentially hold back each group for testing, while training the machine learning model on all the other groups. Subsequently, $R^2$ is computed for each fold by evaluating the model on the withheld test set. In addition to $R^2$, we evaluate model performance using the Mean Absolute Error (MAE) and the Mean Absolute Percentage Error (MAPE) (see~\ref{sec:perfmes}). Note that while $R^2$ is to be maximized, we would like to minimize both MAE and MAPE. \\

For each of these metrics, averages over the 10 cross-validation folds are shown in Table~\ref{tab:cross1}. The average training time is also given. From these results one can see that for data set $\mathbf{A}$, the RF model outperforms all the other models. The $R^2$ of the RF model is close to the $R^2$ for the MLP model, but it has a lower $\textsc{MAE}$, which shows that its average error is of a lower magnitude. For data set $\mathbf{B}$, the MLP shows the highest performance, and for data set $\mathbf{C}$ the CNN has the highest value for all recorded metrics. As can be expected from the experiments in Section~\ref{sec:hyp}, the performance of the PEN is rather poor compared to the other models. Additionally, it was not able perform a fitting operation on the complete data set $\mathbf{B}$, since the polynomial expansion of the input data causes the machine to go out of memory. Note also that the fitting times for both neural network models (MLP and CNN) are significantly higher than for the other models but show very good performance at the cost of an increased learning effort. Notably, all models (except for PEN) show the lowest error on data set$~\mathbf{C}$, which shows that the variance introduced by the LHS is beneficial in training a machine learning model. Overall, we can observe that the models (except for PEN) show exceptionally high $R^2$ scores on all data sets, which means that they can emulate the behaviour of the ANF model almost perfectly with minimal deviation. \\

\begin{table}[tb] 
    \centering
\begin{tabular}{c|c|c|c|c|c|c}
                     & \textbf{Metric} & \textbf{PEN} & \textbf{MLP}   & \textbf{CNN}   & \textbf{GB}     & \textbf{RF}    \\ \hline
\multirow{4}{*}{$A$} & $R^2$           & 0.599        & 0.979          & 0.942          & 0.940           & \textbf{0.980} \\
                     & $\textsc{MAE}$  & 0.950        & 0.177          & 0.312          & 0.304           & \textbf{0.137} \\
                     & $\textsc{MAPE}$ & 0.277        & 0.045          & 0.079          & 0.075           & \textbf{0.034} \\
                     & time (s)        & 783.85       & 24419.65       & 4830.11        & \textbf{43.23}  & 202.56         \\ \hline
\multirow{4}{*}{$B$} & $R^2$           & X            & \textbf{0.998} & 0.987          & 0.995           & 0.983          \\
                     & $\textsc{MAE}$  & X            & \textbf{0.196} & 0.551          & 0.199           & 0.353          \\
                     & $\textsc{MAPE}$ & X            & 0.035          & 0.116          & \textbf{0.020}  & 0.036          \\
                     & time (s)        & X            & 227128.67      & 232049.12      & \textbf{905.37} & 8819.14        \\ \hline 
\multirow{4}{*}{$C$} & $R^2$           & 0.197        & 0.997          & \textbf{0.999} & 0.998           & 0.998          \\
                     & $\textsc{MAE}$  & 4.870        & 0.188          & \textbf{0.147} & 0.272           & 0.310          \\
                     & $\textsc{MAPE}$ & 2.112        & 0.064          & \textbf{0.061} & 0.119           & 0.142          \\
                     & time (s)        & 6518.32      & 302.36         & 956.77         & \textbf{37.53}  & 137.46         \\ \hline
\end{tabular}
    \caption{Results for 10-fold cross-validation for all data sets and machine learning models. Average over cross-validation folds for prediction quality metrics $R^2, \textsc{MAE}$ and $\textsc{MAPE}$ are shown, in addition to the average fitting time in seconds. Best values attained for each experiment set are shown in boldface. Results of PEN on data set $\mathbf{B}$ are dropped due to the prohibitively large size of the data set.}
    \label{tab:cross1}
\end{table}

\begin{table}[tb]
    \centering
\begin{tabular}{c|c|c|c|c|c|l}
                                                                & \textbf{Metric} & \textbf{PEN} & \textbf{\textbf{MLP}}   & \textbf{\textbf{CNN}}   & \textbf{GB}     & \textbf{\textbf{RF}}  \\ \hline

\multirow{4}{*}{$\mathbf{B}_{grouped}$}                         & $R^2$           & X            & 0.665                   & \textbf{\textbf{0.963}} & 0.920           & 0.824                 \\
                                                                & $\textsc{MAE}$  & X            & \textbf{\textbf{0.752}} & 0.805                   & 0.926           & 1.433                 \\
                                                                & $\textsc{MAPE}$ & X            & 0.186                   & 0.138                   & \textbf{0.093}  & 0.152                 \\
                                                                & time (s)        & X            & 178285.10               & 205791.83               & \textbf{626.32} & 7632.34               \\ 
\hline
\multirow{4}{*}{$\mathbf{C} \rightarrow \mathbf{B}_{f_{1213}}$} & $R^2$           & -0.991       & 0.954                   & \textbf{\textbf{0.974}} & 0.692           & 0.366                 \\
                                                                & $\textsc{MAE}$  & 1.724        & 0.176                   & \textbf{\textbf{0.114}} & 0.497           & 0.656                 \\
                                                                & $\textsc{MAPE}$ & 2.272        & 0.215                   & \textbf{\textbf{0.147}} & 0.521           & 0.721                 \\
                                                                & time (s)        & 179.01       & 383.65                  & 3181.94                 & \textbf{43.99}  & 148.52                \\
\hline
\end{tabular}

    \caption{Results for the Leave-P-Groups-Out with data set $\mathbf{B}$, and the cross-prediction (predicting data set $\mathbf{B}$ with a model trained exclusively on data set $\mathbf{C}$) experiments. Prediction quality metrics $R^2, \textsc{MAE}$ and $\textsc{MAPE}$ are shown, in addition to the fitting time in seconds. Best values attained for each experiment set are shown in boldface. Note that the top row shows averages over a 10-fold cross-validation, while the bottom row only shows a single experiment.}
    \label{tab:cross2}
\end{table}

\paragraph{Leave-P-Groups-Out cross-validation}
The fact that data set $\mathbf{B}$ contains data for 1\,296 unique stimulus waveform shapes, allows us to do an additional experiment where we can test a model's quality when predicting unseen waveforms. In the previous experiment, this was not explicitly enforced, and thus the machine learning models could potentially still learn from similar data from the same waveform shape. This can be enforced by splitting the data into groups, where each group contains the complete set of data for one stimulus waveform, and then perform a \textit{Leave-P-Groups-Out} cross-validation experiment. In a similar fashion as with the previous 10-fold cross-validation experiment, we again create 10 cross-validation folds, but now the folds are generated by using random subsets of groups, and not of individual samples. This ensures that every validation set contains only data from groups that were not used to train the model. Since we group by waveform, we can thus use this experiment to answer the question: \textit{If we generate a model using a set of known stimulus waveforms, how well can we expect the model to generalize to new waveforms?} \\

The top row of Table~\ref{tab:cross2} shows the results for this experiment. From these results, we can observe a significant drop in model quality for all models, which is to be expected, due to the increased complexity of the experiment. However, with the RF model, this is more pronounced, which is probably due to the fact that RF models are known to fall short in extrapolation, when predicting samples out of the domain of the training data~\citep{extrarf}. Interestingly, all three other models are showing the best value for one of the metrics, with the highest $R^2$ (.963) attained by the CNN model.

\paragraph{Cross-prediction}
Moreover, since data set $\mathbf{B}$ contains 1\,296 samples for fiber 1213, and data set $\mathbf{C}$ contains \textit{only} samples for fiber 1213, we can use a model trained on data set $\mathbf{C}$ to predict the samples for fiber 1213 from data set $\mathbf{B}$. As we only have a single training ($\mathbf{C}$) and validation set ($\mathbf{B}_{f1213}$), we only do one experiment for each model, and perform no additional cross-validation. With this we can ask the question: \textit{Can we use a model trained on exclusively pseudo-random samples to predict data for waveforms used in clinical practice?} \\

Results for this experiment is shown in the bottom row of Table~\ref{tab:cross2}. From this, we can see that only the neural network (NN) based models are giving satisfactory performance in this experiment, with the CNN clearly outperforming the MLP model. Moreover, both models actually are showing better performance in this experiment than on the Leave-P-Groups-Out experiment. This gives rise to the notion that for training a neural network model (i.e. MLP or CNN) that generalizes to unseen data, the LHS based data set $\mathbf{C}$ is better than data set $\mathbf{B}$. Furthermore, the fact that the other models are not able to do well in this experiment, might mean that they are overfitting to the training data in the previous experiments, which was much more homogeneous with the test data than in this experiment. \\

Overall, the validation experiments presented in the previous paragraphs show that when only predicting samples from single data set source, all models except for the PEN model show satisfactory performance. The GB model is quickest to train, but can be of lower quality than some of the other models tested. When testing for generalizability and evaluating how the models perform on data from other data set sources, we find that the CNN model shows the most promise. \textit{Thus, in the remainder of the paper, we will use the CNN model for our experiments.}

\begin{figure}
    \centering
    \includegraphics[width=\textwidth]{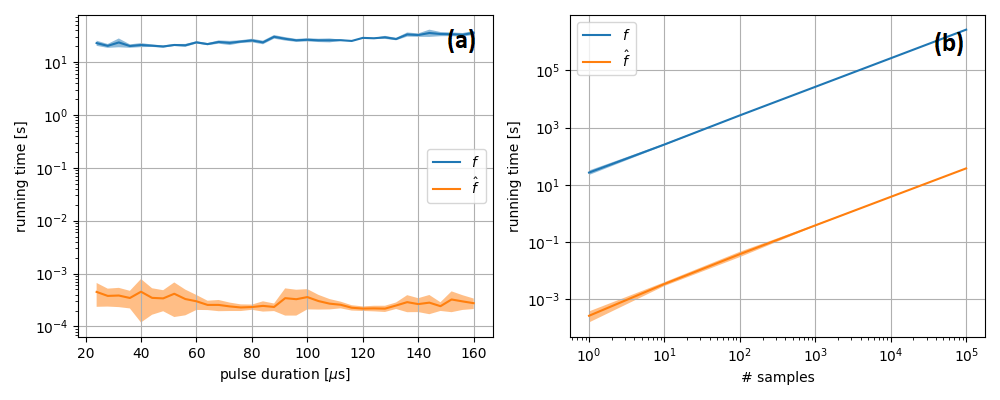}
    \caption{$\log_{10}$ scaled running time in seconds of the both the ANF model $\mathbf{f}$ and the CNN Surrogate $\hat{\mathbf{f}}$ vs. the duration of the pulse \textbf{(a)} and the number of evaluated samples \textbf{(b)}. Samples of running time are generated using bootstrapping, and the average values are shown. The bands represent the standard deviation.}
    \label{fig:rt}
\end{figure}

\paragraph{Running time}
In Figure~\ref{fig:rt} the running (prediction) time of the surrogate model is shown in comparison to the running time of the original ANF model. For the surrogate, the CNN model is shown, since this model showed the best performance in the previous experiments. From this figure, we observe that the surrogate model realizes a speedup of 5 orders of magnitude w.r.t the running time of the ANF model. In addition, we can see that while the running time increases linearly with the duration of the pulse for $\mathbf{f}$ it remains constant for $\hat{\mathbf{f}}$. 

\section{Waveform Optimization}
\label{sec:optimization}
The experiments in this section are based on the methodology presented in~\citet{ga2010}, where a Genetic Algorithm (GA) is used to find energy-optimal stimulus waveforms. As an analytical solution for an energy-optimal stimulus waveform is not easily derived due to the complex non-linear nature of the nerve model, we use heuristic optimization with a real valued GA. \\

A Genetic Algorithm~\citep{ga} is a nature-inspired metaheuristic, which seeks to evolve a population of candidate solutions to an optimization problem through an iterative procedure. By combining the properties of good solutions with small random changes, the GA can move towards better solutions over time. It ranks solutions based on an objective or fitness function $\mathcal{F}(\mathbf{s}) \mapsto \mathbb{R}$, which computes the quality of a candidate solution $\mathbf{s}$ relative to optimization problem the GA is solving. Historically, the GA is a method for solving discrete optimization problems, and the modifications made in~\citet{ga2010} for solving a continuous optimization problem, cause it to be closer to an Evolution Strategy~\citep{es} (ES) than a GA. Therefore, in the remainder of the paper, we will refer to the GA used by~\citet{ga2010} by the overarching term Evolutionary Algorithm (EA) for correctness. \\

Instead of a (computationally expensive) ANF model, we make use of use the surrogate model introduced earlier in this paper to compute the nerve response, which saves a significant amount of computation time. For the surrogate, we will use the CNN model which was re-trained on data set all data from $\mathbf{C} \cup \mathbf{B}_{f1213}$ (see Section~\ref{sec:crossval}). With this experiment, we aim to demonstrate the applicability of the surrogate model as a stand-in replacement of the ANF model, and ask the question: \textit{Can the shape of the stimulus waveform used by the original ANF model be accurately optimized while performing the expensive simulations with the surrogate model?}

\subsection{Experimental setup}
\label{sec:wavesetup}
\begin{figure}[tb]
\centering
\resizebox{0.9\textwidth}{!}{
\begin{tikzpicture}[node distance=4cm]
\tikzstyle{comp2} = [draw, rectangle, rounded corners, minimum height=1cm, minimum width=3cm]

\node (ga) [comp2, fill=lightgray]{\textbf{Evolutionary Algorithm}};
\coordinate[below of=ga] (c);
\node (wv) [comp2,  left of=c, yshift=1.5cm]{$\wave'$};
\node (en) [comp2,  right of=wv, yshift=1cm]{$E(\wave')$};
\node (ap) [comp2,  right of=wv, yshift=-1cm]{$\hat{\mathbf{f}}(\wave') > 1$};
\node (f) [comp2,  right of=ap, yshift=1cm]{$\mathcal{F}(\wave')$};

\coordinate[left of=ga] (c1);
\coordinate[right of=ga] (c2);
\draw [] (ga) --  (c1);
\draw [arrow] (c1) -- (wv);
\draw [arrow] (wv) -- (en);
\draw [arrow] (wv) -- (ap);

\draw [arrow] (en) -- (f);
\draw [arrow] (ap) -- (f);
\draw [] (f) --  (c2);
\draw [arrow] (c2) -- (ga);
\end{tikzpicture}
}
\caption{Schematic overview of the waveform optimization process with a Evolutionary Algorithm. The EA generates a new candidate solution $\wave'$. Energy and the threshold for $\wave'$ (calculated by the surrogate ANF model $\hat{\mathbf{f}}$), are then combined into $\mathcal{F}(\wave')$, which is fed back into the EA.}
\label{fig:sdia2}
\end{figure}
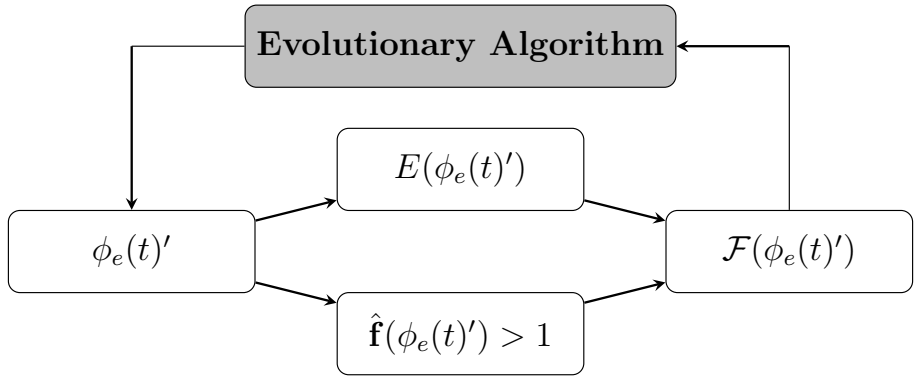

In this work, candidate solutions are stimulus waveforms ($\mathbf{s}=\wave$), which are vectors in $\mathbb{R}^{d}$, where each component represents the amplitude of the waveform at a given time step ($dt$ = $2\mu s$). The fitness of a candidate solution (i.e., how good a candidate solution is) is based on the signal energy of the pulse $E(\wave)$, defined by the sum of the squared magnitude of the waveform signal:
\begin{equation}
\label{eq:ener}
    E(\wave) = dt\sum_{t=0}^{\infty} \vert  \wave \vert^2,
\end{equation}
where $dt$ is the size of the time step. Fitness is defined as $E(\wave)$ plus a penalty term $P$: 
\begin{equation}
    \mathcal{F}(\wave) = E(\wave) + P. 
\end{equation} 
For each wave $\wave$, the threshold value $\thres$ is calculated with the surrogate model $\hat{\mathbf{f}}$. If $\wave$ is stimulating below threshold (i.e. $\thres > 1$) the penalty term $P$ is assigned to 0.01, which is several orders of magnitude greater than $E(\wave)$, and 0 otherwise:
\[
    P = 
\begin{dcases}
    1e5, &  \thres > 1\\
    0,       & \text{otherwise}.
\end{dcases}
\]
A schematic overview of this process is shown in Figure~\ref{fig:sdia2}. \\

We conduct experiments for the pulse durations:  $\{20, 40,\dots, 140, 160\}$ $\mu$s, which are encoded in the EA by real valued vectors of dimensionality: $\{10, 20,\dots, 70, 80\}$. In order to satisfy the constraint in Eq.(~\ref{eq:const}), we investigate the following strategies:
\begin{enumerate}
    \item \textbf{Projection:} In a similar fashion as was done in Section~\ref{sec:lhs}, candidate solutions are projected onto a non-violating hyper-plane. This constricts the search space of the optimization algorithm to $d-1$ but allows for non-biphasic pulses to be generated. 
    \item \textbf{Mirroring:} Only the (first) positive phase of the stimulus is optimized, and the second phase consists of the mirror opposite with respect to zero of the first phase to balance the charge.  
    \item \textbf{Balancing:} Only the positive phase of the stimulus is optimized, and a rectangular pulse is used to balance the charge of the first phase. For the balancing rectangular wave, we investigate both preceding and ensuing pulses (referred to later here as `balancing\_pre' and `balancing\_post'). 
\end{enumerate}

Note that for both balancing and mirroring, the optimization problem becomes considerably simpler, since instead of $d$ points, only $\frac{d}{2}$ points have to be optimized by the optimization algorithm.  \\

The EA was implemented to the specifications given in~\citet{ga2010}, with two-point crossover and where mutation of each candidate solution occurs via scaling with a random factor: $\mathbf{s}_i^{t} = \mathbf{s}_i^{t-1} \cdot \mathcal{N}(1, 0.025)$. For the projection constraint handling strategy, a scaling mutation operator would constrict pulses to the same polarity, which is why for this strategy, the mutation occurs not via scaling, but by shifting with a random factor drawn from a normal distribution, i.e: $\mathbf{s}_i^{t} = \mathbf{s}_i^{t-1} + \mathcal{N}(0, 0.025)$. A population size of 50 was used, where for each generation the top 10 candidate solutions were passed on to the next generation (elitism). For every combination of the aforementioned constraint handling strategies and each dimensionality, the EA was run with 10 independent trails for 10\,000 generations. For every trial, the EA thus conducts a total of $400\,010$ evaluations of the surrogate. \\ 

\subsection{Results}
\label{sec:waveres}

\begin{figure}
    \centering
    \includegraphics[width=\textwidth]{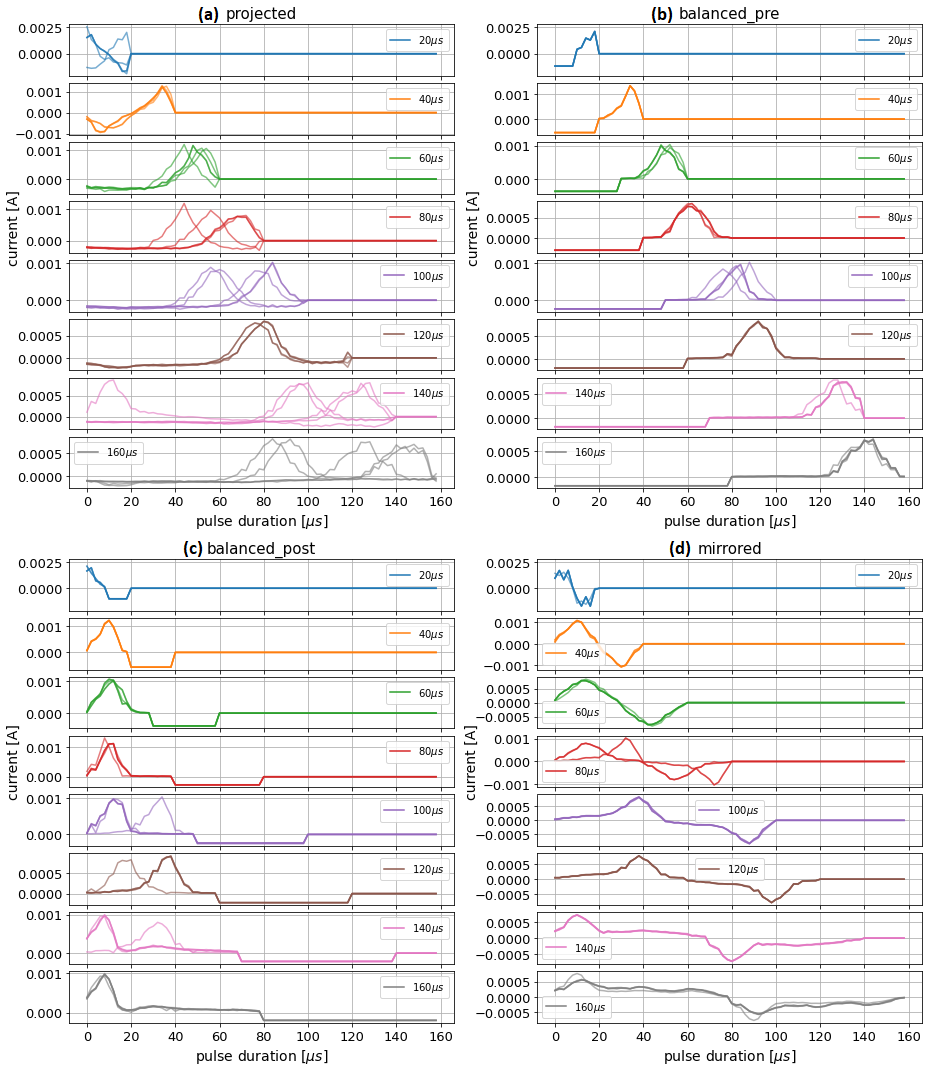}
    \caption{The optimal EA waveforms, grouped by pulse duration and constraint handling strategy. Each curve represents a best solution given by the EA for a trial, and is scaled by $\thres$ computed by the ANF model. For each sub graph, 10 trials are shown. Figure \textbf{(a)} shows the waveforms generated for the projected constraint handling strategy and Figures \textbf{(b-d)} resp. show the waveforms generated by the balanced\_pre, balanced\_post and mirrored strategies.}
    \label{fig:ga_waveforms}
\end{figure}

Looking at the shape of the pulses generated by the EA (see Figure~\ref{fig:ga_waveforms}), some interesting observations can be made. For a pulse duration of $> 40\mu$s almost all the generated pulses have the positive phase resembling a Gaussian curve and an elongated balancing phase, which is in line with the curves found in~\citet{ga2010}. The exception to this is the mirrored strategy, where due to the enforced symmetry both phases are resembling Gaussians. The shorter pulses have peaks that are notably sharper, and are exponential in shape for some cases (e.g. 40$\mu$s balanced\_pre). The results are quite stable across trials, and we can see most of the trial runs are all converging to similar shapes. This indicates that a (local) optima is being found by the EA, and we can confidently solve the optimization problem. The least stable results are from the projection constraint handling strategy, but often only the peak offset is varying. As was mentioned before, the optimization task when using the projection constraint handling strategy is significantly more complex, due to the greater number of variables optimized. With this in mind, it might be that strategy needs a longer time to converge, and hence the less stable results. This increased complexity, however, comes with the added benefit that pulses generated by this strategy are not constricted to biphasic pulses. Interestingly however, the shapes this strategy generates, are still of that general shape, with only a few exceptions of tri-phasic pulses (e.g. for 120$\mu$s). Moreover, the strategy seems to converge into shapes that are almost mono-phasic, where for longer pulse durations, the elongation is only used to balance the negative charge more evenly over a longer period. By definition, such shapes are also being generated by the balancing strategy, which shows that even though the balancing strategies are restricted to a biphasic shape, the shapes they converge to are close to the converged shapes of an unrestricted strategy. This means that the increased flexibility that is introduced by the projection strategy is not required for this specific problem, since the shapes that it converges to can also be generated with a simpler strategy that uses balancing. Nonetheless, when comparing the pulses of the balancing pre and post strategies, we see that the projection strategy is actually able to ultimately find shapes that are similar to that of the better (lower energy) of these two strategies (e.g. 20 $\mu$s).

\subsubsection{Validation}
\label{sec:validation}
In order to validate the predictions made by the surrogate $\hat{\mathbf{f}}$ during the optimization process, we also compute the threshold value with the ANF model. We use this to compute a validated energy value for each of the waveforms that resulted from the optimization, by scaling each of these waveforms by the threshold computed by the original ANF model $\mathbf{f}(\wave)$:
\begin{equation}
    E^*(\wave) = E(\mathbf{f}(\wave)\wave)
\end{equation}
In results presented in the sequel, we refer to this validated energy value when discussing energy, and objective function value $\hat{\mathbf{f}}$ otherwise. \\

\begin{figure}[tb] 
    \centering
    \includegraphics[width=\textwidth]{ga_objective_values_per_dim_mean_with_pre.png}
    \caption{\textbf{(a)} pulse duration vs. the value of $\mathcal{F}$ of the best solutions found by the EA computed using the surrogate, grouped by constraint handling strategy. \textbf{(b)} pulse duration vs. energy, scaled with $\thres$ computed by the ANF model. \textbf{(c)} pulse duration vs. energy values, and the EA-generated waveforms (aggregated over all constraint handling strategies) are compared against the energy values of the predefined pulse shapes (see Section~\ref{sec:stim}). Each line with dots shows the mean value over 10 trials, with the shaded area as standard deviation.}
    \label{fig:dim_vs_objective} 
\end{figure}

\subsubsection{Energy efficiency}
The objective function (left) and energy (right) values for the final solutions provided by the EA for each constraint handling strategy are shown in Figure~\ref{fig:dim_vs_objective}. When comparing Figure~\ref{fig:dim_vs_objective}\textbf{(a)} with the Figure~\ref{fig:dim_vs_objective}\textbf{(b)}, we can observe a discrepancy between the actual energy value and the predicted value by the surrogate, and on average, the surrogate predicts a 6.61\% lower threshold. This seems to be especially prevalent in the waveforms generated using the mirrored strategy, for which the surrogate tends to predict a lower threshold than the ANF model. \\

Figure~\ref{fig:dim_vs_objective} shows that, in general, the energy of the solutions decreases with pulse duration. The waveforms generated by balancing\_pre and projection strategies are yielding the most energy efficient pulses, while the mirroring strategy is less efficient. Balancing with a rectangular phase after the optimized phase is also resulting in pulses with lower energy than balancing before the optimized phase. Figure~\ref{fig:dim_vs_objective} \textbf{(c)} shows the union of all the constraint handling strategies as a single line in comparison with the energy values for the predefined pulses from Section~\ref{sec:stim}. From this we can observe that for all pulse durations, the optimized waveforms are of \textit{lower energy}, and the EA provides better solutions than the handcrafted pulses (data set $\mathbf{B}$). This effect becomes more even pronounced with a longer pulse duration.  \\

\begin{figure}[tb] 
    \centering
    \includegraphics[width=\textwidth]{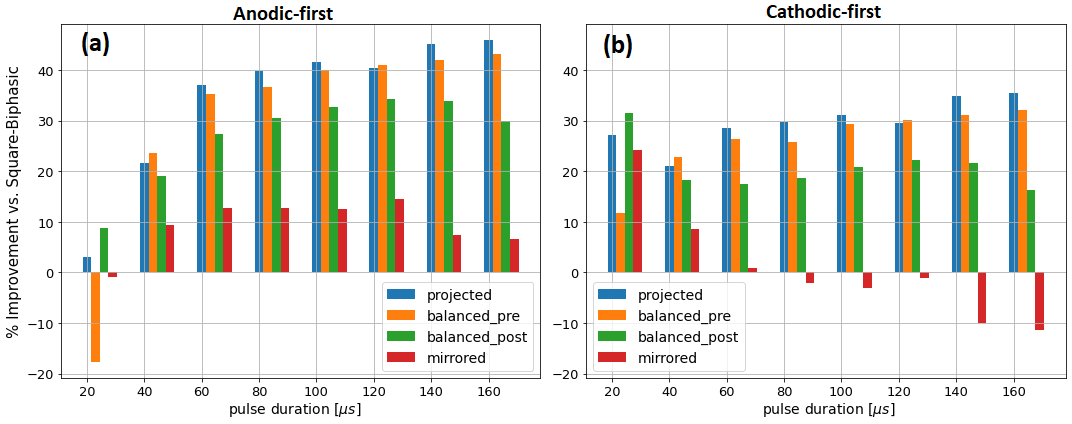}
    \caption{The figure shows dimension vs. the energy value of the best solutions found by the EA in a series of 10 runs grouped by constraint handling strategy, relative to the energy of a square biphasic pulse. \textbf{(a)} shows the comparison with an anodic-first square wave, \textbf{(b)} shows the comparison with a cathodic-first square wave. Energy values are computed using the ANF model.}
    \label{fig:objective_vs_square}
\end{figure}

Figure~\ref{fig:objective_vs_square} shows a comparison between the energy of the pulses generated by the EA and the energy of the typically used square wave of that same pulse duration, for both anodic-first and cathodic-first pulses. For each strategy, the solution with the minimal energy over the 10 trials is shown. From this figure, we can observe a similar pattern as in Figure~\ref{fig:dim_vs_objective}, in that the energy efficiency relative to a square wave tends to increase with pulse duration, even though the trend is not as pronounced. Interestingly, it seems that for shorter pulse lengths, anodic square pulses seem more efficient, while for longer pulses this patterns reverses, which seems to be contrasting previous works~\citep{macherey2008higher, undurraga2010polarity, kalkman2022}.  For shorter pulse lengths ($< 20\mu$s), the anodic-first square wave is already quite energy efficient, and only the balanced\_post strategy can yield a significantly more energy efficient pulse. Overall, we can observe that the waveforms generated by the EA are of a substantially lower energy over all pulse durations ($> 20\mu$s), and on average have 31\% lower energy than a cathodic-first, and 41\% lower energy than a anodic-first square wave of that same pulse duration. \\

\section{Conclusion} 
\label{sec:conclusion}
In this work we have presented a surrogate to the ANF model from~\citet{kalkman2022} and used it to conduct an optimization experiment. The surrogate model was able to predict the behavior of the ANF model with very high confidence, whilst reducing computation time by five orders of magnitude. Five different machine learning models were tested, of which the CNN model was the most promising and best able to extrapolate its prediction to unseen experimental conditions. Our data was comprised of a combination of LHS samples and handcrafted simulation data, and we found that the CNN model was able to accurately predict the handcrafted simulation data while having only trained on the (pseudo-random) LHS samples. With this study, we have aimed to show that through the use of a surrogate a complex process such as the ANF model can be accurately approximated, allowing for large scale experimentation. This is not restricted to this specific ANF model however, and this methodology \textit{can be extended} to other ANF models and processes. For example, the prediction of the time to an action potential or the detection of the node where the action potential starts. \\

The second part of our study used the CNN surrogate model with an EA to optimize stimulus waveforms for energy efficiency at threshold level. Our research suggests an alternate waveform to the commonly used square biphasic pulses. The waveform resembles a positive Gaussian curve offsetted by an elongated rectangular negative phase, which is in line with earlier research~\citep{sahin2007non, ga2010}. The shape is in disagreement with the pulses generated in~\citet{ciga}, where decaying exponential waveforms were found to be the most efficient, but this might be due to the fact that the ANF model used here includes a 3D model interface, and we applied fewer restrictions on waveform shape. The shapes differ with varying pulse duration, and we found that the energy efficiency of the waveforms generated by the EA increases with pulse duration. We observed 8\% - 45\% energy decreases when comparing to a square biphasic pulse for total pulse durations of 20 - 160 $\mu$s. Additionally, we found that for shorter pulses, an anodic-first square wave is more energy efficient, while for longer pulses a cathodic-first pulse is better. The generated waveforms were also considerably more energy efficient than those from a large set of handcrafted pulses. \\

We propose a flexible approach that uses projection for generating charge-balanced pulses, which can consistently produce the most energy efficient pulses for each of the charge-balancing strategies tested. Interestingly, the shapes generated with this unrestricted strategy are close to the converged shapes of a restricted strategy, which was used in earlier research. This indicates that even while considering the full possible range of waveform shapes, the most energy optimal shape is still a relatively uncomplicated biphasic waveform.  \\

It must be noted that, even though our modelling \textit{in silico} suggests these potential energy savings, further work with in patient testing is required to validate these results. This would then also consider the effects of capacity and the power dissipated at the electrode-tissue interface to the realized energy savings. Moreover, state-of-the-art analysis shows that the effect of an adaptive power supply can have a drastic impact on the energy efficiency of certain waveform shapes~\citep{delft2022}, which is not taken into account here. Future work might focus on a wider range of pulse durations, or the extension of this approach to other ANF models. Additionally, the benefits of using a surrogate modelling approach to full speech simulation might also be an interesting line of research. \\

\section{Acknowledgements}
We would like to thank Randy Kalkman for his vital support in collecting the simulation data, and Hao Wang for his advice and feedback. This work was supported by Health Holland, under research project 'TEMPORAL' in collaboration with Advanced Bionics.

\bibliography{main} 

\newpage
\appendix
\section{Potential distributions}
\label{sec:pot}
\begin{figure}[h] 
    \centering
    \includegraphics[width=\textwidth]{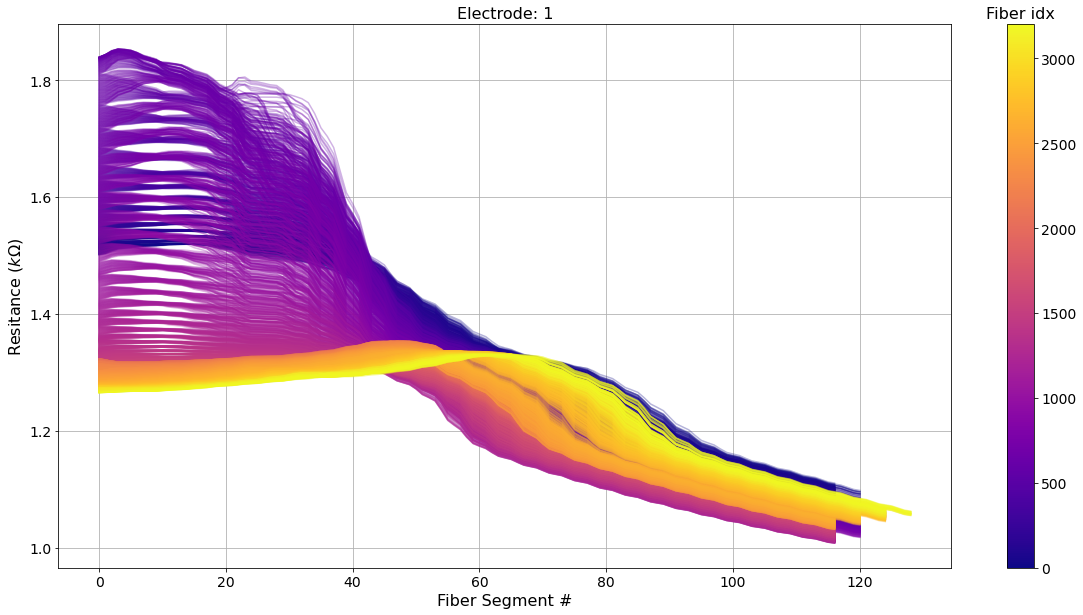}
    \caption{Electric potential distributions of 3200 Healthy fibers induced by electrode 1 taken from HC3A cochlear geometry, with PM electrode placement. The hue of a line represents the id of fiber, which are ordered from base to apex.}
    \label{fig:bpp}
\end{figure}

As mentioned in Section~\ref{sec:volc}, the potential distribution $\pot$ is a set of $\approx 115$ real-valued measurements computed along a fiber $f$ of potential energy induced by an electrode $e$, measured at evenly spaced segments. The fact that not every fiber in the model has the same length, and thus not the same number of segments, causes the number of measurements to differ from fiber to fiber. An example of the potential distributions, induced by electrode 1 on 3\,200 fibers modeled using the HC3A~\citep{Kalkman2015a} cochlear geometry is shown in Figure~\ref{fig:bpp}. As electrode numbers are ordered from base to apex, meaning this electrode is placed closest to the base of the cochlea, we can see that fibers located at the base (darker colors) show greater response to this electrode than fibers located more deeply in the cochlea. \\

Additionally, we observe from the figure that for each fiber the potential distribution follows the same curve, only the magnitude differs. From this observation it follows that we can well approximate each of these curves by a polynomial of degree $k$: 
\begin{equation}
    f_w(x) = \sum_{i=0}^k w_i x^i,    
\end{equation}
parameterized by a set of real-valued weights $w$, obtained using e.g. least squares regression. By using the weights from this polynomial as representation of $\pot$, the number of features for this parameter-component is effectively reduced and standardized from $\approx 115$ to the polynomial degree $k$. \\

\begin{figure}[tb] 
    \centering
    \includegraphics[width=0.8\textwidth]{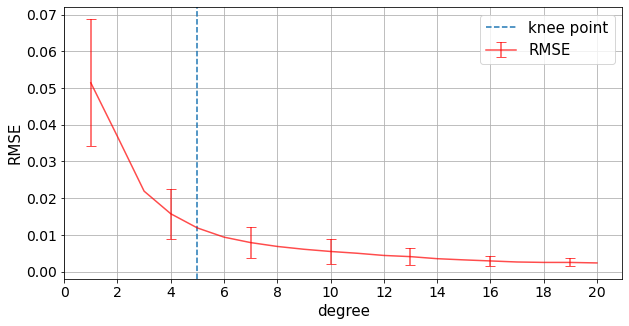}
    \caption{Average RMSE vs. the degree of the fitted polynomial with standard deviation displayed as error bars. The knee-point of the RMSE curve is shown in dashed line.}
    \label{fig:rmse_degree}
\end{figure}

In order to determine the appropriate degree $d$, we compute the Root Mean Squared Error (RMSE) between the original potential distributions and their corresponding polynomial representations for $d$ $\in [1, 20]$ for data set $\mathbf{A}$, which is shown in Figure~\ref{fig:rmse_degree}. Note that the values shown are averages over all fiber and electrode combinations in our data. Subsequently, we compute the knee point of the RMSE vs. degree curve using the \textit{Kneedle}~\citep{kneedle} algorithm. As can be seen in the figure, this results in the value 5, which we use for the degree of polynomial representation of $\pot$. 

\section{Performance Metrics}
\label{sec:perfmes}
\textbf{Mean Absolute Error $(\textsc{MAE})$} or $\mathcal{L}1$-norm, representing the expected value of the absolute error. It is defined on $[\infty, 0]$ as: 
\begin{equation}
    \textsc{MAE}(y, \hat y) = \frac{1}{n} \sum_{i=1}^n \vert y_i - \hat y_i \vert 
\end{equation}
\textbf{Mean Absolute Percentage Error $(\textsc{MAPE})$} represented the expected value of the absolute error of the prediction in proportion to the relative magnitude of the output. This makes the metric more sensitive to relative errors.  It is defined on $[\infty, 0]$ as: 
\begin{equation}
    \textsc{MAPE}(y, \hat y) = \frac{1}{n} \sum_{i=1}^n \frac{\vert y_i - \hat y_i \vert}{\max(\epsilon, \vert y_i \vert)},
\end{equation} where $\epsilon$ is set to an arbitrary small value.

\end{document}